\documentstyle[12pt,epsf]{article}
\newcommand{\bc}{\begin{center}}
\newcommand{\ec}{\end{center}}
\newcommand{\bd}{\begin{displaymath}}
\newcommand{\ed}{\end{displaymath}}
\newcommand{\be}{\begin{equation}}
\newcommand{\ee}{\end{equation}}
\newcommand{\ba}{\begin{eqnarray}}
\newcommand{\ea}{\end{eqnarray}}
\newcommand{\bt}{\begin{tabular}}
\newcommand{\et}{\end{tabular}}

\begin{document}
\begin{flushright}
APCTP--98--21
\end{flushright}

\vspace{0.5cm}

\begin{center}
{\Large\bf
Exclusive and Inclusive Decays of the $B_c$ Meson in\\[5mm]
the Light--Front ISGW Model }\\[1cm]
{\it A.Yu.Anisimov, P.Yu.Kulikov,  I.M.Narodetskii, K.A.Ter--Martirosyan}\\[3mm]
{\small Institute of Theoretical and Experimental Physics, 117259 Moscow,
Russia}\\[2mm]
\end{center}
\vspace{2cm}

\begin{abstract}
\noindent
We investigate the total decay rate of the (ground state) $B_c$ meson within
the framework of the relativistic constituent quark model formulated on
the light-front (LF).
satisfactory.
The {\it exclusive} semileptonic (SL) and nonleptonic
(NL) beauty and charm decays of the $B_c$  meson are described through vector
and axial hadronic form factors, which are calculated in terms of the overlap
of the parent and daughter meson LF wave functions. The latters are derived via the
Hamiltonian LF  formalism using as input the update version of the
ISGW model.
The {\it inclusive} SL and NL partial rates are calculated within a convolution
approach inspired by the partonic model and involving the same $B_c$ wave
function which is used for evaluation of the exclusive modes.
We predict the partial rates for  74    exclusive SL and NL channels
and 43  inclusive
partial rates corresponding to the underlying $\bar{b}\to \bar{c}$ and $c\to
s$ quark decays. Based on our approach we find $\Gamma^{\bar{b}
}(B_c)= 0.52 \pm 0.02 ~ ps^{-1}$, $\Gamma^{c}(B_c)= 0.98 \pm
0.07 ~ps^{-1}$, where the theoretical uncertainty is dominated by the
uncertainty in the choice of the threshold values at which the hadron continuum
starts. For the $B_c$ lifetime we obtain $\tau_{B_c}= 0.63 \pm 0.02~ps$ in a good
agreement with the prediction obtained using the nonrelativistic operator product
expansion.  We also predict decay rates for many specific weak transitions of
$B_c$. In particular, for the branching fractions of the $B^+_c \to
J/\psi\mu^+\nu_{\mu}$,  $B_c^+\to J/\psi\pi^+$ and $B_c^+ \to J/\psi~X $
decays
we obtain $1.7\%$, $0.1 \%$ and $13.2\%$, respectively.
\end{abstract}

\vspace{1cm}
\bc
Accepted for publication in  {\it Physics of Atomic Nuclei}
\ec

\vspace{1cm}


\newpage
\section{Introduction}
Weak decays of hadrons containing a heavy quark present the most direct way
to determine the Cabbibo--
Kobayashi--Maskawa (CKM) matrix and to test our present knowledge
of the QCD confinement scale inside hadrons. Among various heavy flavor
hadrons the $B_c$ meson,
the bound state of the $\bar{b}c$
system with open charm and beauty,
is particularly interesting.
The theoretical interest in the study of
the $B_c$ meson is stimulated by
the experimental search at CDF and LHC. Since last year, ALEPH has a
very clean signal $B^+_c \to J/\psi\mu^+\nu_{\mu}$ with a measured $B_c$
mass $5.96^{+0.25}_{-0.19}~
GeV/c^2$ \cite{ALEPH97}. A new preliminary OPAL analysis \cite{OPAL97}
finds two $J/\psi
\pi^+$ candidates with the masses $6.29 \pm 0.17$ and $6.003 \pm
0.06~GeV/c^2$. Recently the CDF Collaboration  reported the
observation of $B_c$ in $1.8~TeV ~p\bar{p}$ collisions using the CDF detector
at Fermilab Tevatron \cite{CDF98}. The CDF results for the mass and lifetime
are $M_{B_c}=6.40 \pm 0.39(stat) \pm 0.13 (syst)~ GeV/c^2$ 
and $\tau_{B_c}=0.46^{+0.18}_{-0.16}(stat)
\pm 0.03(syst)~ps$.
The physics of $B_c$ mesons has stimulated
much recent works on their properties, weak decays and production cross
section on high energy colliders.
A comprehensive analysis of the
$\bar{b}c$ spectroscopy and strong
and electromagnetic decays of the excited states has been given in \cite
{GKLT95}.

Similarly to $D$ and $B$ mesons,
the ground $\bar{b}c$ state is stable against strong or electromagnetic
decay due to its flavor content and disintegrates only via weak interactions.
The weak $B_c$ meson decays occur mainly through the CKM favored $b \to{c}$
transitions with $c$ being a
spectator, and $c \to s$ transitions with $\bar{b}$ being a
spectator. Weak decay
properties of the ground state $\bar{b}c$ including semileptonic (SL) and
various exclusive nonleptonic (NL) modes have already been discussed
in refs. \cite{DW89}--\cite{CC94}.
In refs. \cite{LM91}, \cite{CC94} the $B_c$ lifetime has been
estimated on the basis
of a modified spectator model, where the phase space for the free quark decay
is modified to account for the physically accessible kinematical region
\cite{LM91}, or the $\bar{b}$ and $c$ quark masses are reduced by the
binding energy
to incorporate bound state effects \cite{EQ94}. An estimation of
the $B_c$ lifetime $\tau_{B_c}$
using a modified spectator
model and information gained from the calculation of dominant
exclusive modes is given in \cite{Q93}.
A careful analysis of the $B_c$ lifetime, performed using the
nonrelativistic QCD and including the nonrelativistic corrections up to
$v^2/c^2$, has been carried in \cite{BB96}, see also \cite{B96}. The wide
range of lifetimes $\tau_{B_c}$ reported in these papers, reflects the
uncertainty
due to the various model
assumptions on the modification of the free decay rates due to the bound state
effects and the limited knowledge of the heavy quark masses.

Weak decays of charmed and bottom hadrons are particularly simple in the
limit of infinite heavy quark mass, where the decay rate of a hadron $H_Q$,
containing a heavy quark $Q$ is completely determined by the decay rate of
the heavy quark itself. In this limit, one might expect that $\Gamma(B_c)
\approx \Gamma(\bar B^0)+\Gamma (D^0)$. If this result were hold, the $B_c$
lifetime would be rather short, namely $\tau_{B_c} \approx 0.3~ps$, and $B_c$
decays would be dominated by $ c\to s$ decay over $\bar b \to \bar c$ decay
in the ratio of
roughly $4:1$. In reality, heavy hadrons are the bound states of heavy quarks
with light constituents. The inclusion of these soft degrees of freedom
generates important contributions due to the preasymptotic effects,
like the binding effects and the Fermi motion of a heavy quark inside the
hadron. These effects have a significant impact on the lifetime and
SL branching ratios.
The leading nonperturbative effect is described by a distribution
function $F(x)$,
("shape function"), which arises as a result of the resummation of
the heavy quark expansion \cite{SF} and has been also incorporated into
phenomenological models of inclusive decays, first in refs.
\cite{AP79}, \cite{ACCMM82},
and lately in refs. \cite{JP94}--\cite{GNST97}. The actual calculation
of $F(x)$ is a difficult nonperturbative problem which in practice
introduces considerable uncertainties in the evaluation of the hadron
lifetime. An important advantage of the $b\bar{c}$ system is the applicability
of a quark potential model treatment \cite{GKLT95}, \cite{EQ94}, \cite{IS95}.
In what follows we assume that, instead of QCD with its complicated
dynamics of infinite number of degrees
of freedom in the light cloud, we consider a constituent bound--state
problem of a heavy
quark interacting with a lighter one {\it via} a potential.
Then, using the formalism of the LF relativistic quantum mechanics
\cite{GNS96}, \cite{DKND97} it is
possible to encode all the nonperurbative
QCD effects in a LF quark model wave function
$\psi(x,p^2_{\bot})$
of a heavy hadron. The internal motion of a heavy
$Q$--quark inside the heavy flavor meson is described by the distribution
function $|\psi(x,p^2_{\bot})|^2$, which represents the
probability to find a heavy quark
carrying a LF fraction $x=p_Q^+/P_{H_Q}^+$ of the meson momentum and a
transverse relative momentum squared
$p_{\bot}^2={\bf p}_{\bot}^2$.

A priory, there is no connection between equal--time (ET) wave function
$w(k^2)$ of a constituent quark model and LF wave
function $\psi(x,p^2_{\bot})$. The former depends on the center--of--mass
momentum squared $k^2=|{\bf k}|^2$, while the latters depends on the LF
variables $x$ and $p_{\bot}^2$.
However, there is a simple operational
connection between ET and LF wave functions \cite{C92}. The idea is to find a
mapping between the variables of the wave functions that will turn
a normalized solution of the ET equation of motion into a normalized solution
of the different looking LF equation of motion. That will allows us to
convert the ET wave function, and all the labor behind it, into a usable
LF wave function. This procedure amounts to a series of
reasonable (but naive) guesses about what the solution of a relativistic
theory involving confining interactions might look like.

We convert from ET to LF momenta by leaving the transverse
momenta unchanged, $ {\bf p_{\bot}}={\bf k_{\bot}} $, and letting
\be
\label{1.1}
k_z=(x-\frac{1}{2})M_0+\frac{m_{sp}^2-m_Q^2}{2M_0},
\ee
where $M_0=\sqrt{k^2+m_Q^2}+\sqrt{k^2+m_{sp}^2}$ is the free mass
operator with $m_{sp}$ being the mass of the quark--spectator. Now
we obtain the LF wave function from
\be
\label{1.2}
\psi(x,p_{\bot}^2)
=\sqrt{\frac{M_0}{4x(1-x)}\cdot
\left[[1-\left(\frac{m_Q^2-m_{sp}^2}{M_0^2}\right)^2\right]}
\cdot \frac{w(k^2)}{\sqrt{4\pi}}.
\ee
It can be easily verified that
$\int\limits^1_0dx\int d^2p_{\bot}|\psi(x,p_{\bot}^2)|^2=1$ provided
$\int d^3k |w(k^2)|^2=1$. It is wave functions made kinematically
relativistic in this fashion, that were used to calculate the form
factors of heavy--to--heavy and heavy--to--light exclusive transitions in
refs. \cite{GNS96}, \cite{DKND97}.

A relevant feature of our approach is that both exclusive and inclusive
decays are consistently treated in
terms of the same heavy quark wave function $\psi(x,p^2_{\bot})$.
So far, this approach has been applied only to the exclusive and inclusive
partial widths of
$\bar{B}^0$ \cite{GNST97}, where it has been found that the overall
picture is quantitatively satisfactory. In this paper we extend previous
calculations
to compute the lifetime
and various decay branching fractions of $B_c^+$.

The paper is organized as follows. In Section 2, after a general overview
of the model, we set the framework for our theoretical calculation of
heavy meson lifetimes in the LF quark constituent model. Here we present
general analysis focusing on calculation of inclusive SL and NL
decay rates.
In Section 3 we apply the model to calculate various
beauty and charm decay rates of the $B_c$ meson in parallel with those of the
beauty decays of the $\bar{B}^0$ meson. In the former case we include both
weak annihilation and Pauli interference contributions to the total width.
Section 4 completes the paper with a summary and conclusions. Technical details
of the calculation are given in Appendix.

\section{Description of the model}
Without going into too many details, we present first a short outline
of the method of calculation of heavy meson partial widths, which is discussed
at length in \cite{GNST97}. We introduce the necessary
definitions and describe the main steps in the calculation of exclusive and inclusive
decay widths of the heavy mesons.
We will make several strong
assumptions  to obtain a model as simple as possible. Yet, we will check
that it agrees, within reasonable limits, with available data on
the SL and NL $\bar{B^0}$--meson decays, as well as with other theoretical
predictions.
We thus feel confident that this model can be used advantageously to
obtain rough estimation for other hadrons, such as $B_c$.

\subsection{Kinematics}
Consider the SL decay rates first. Instead of considering
the exclusive modes individually we will sum (in the representative case
of $\bar{B}^0$ decay) over all possible charmed final
states $X_{\bar{c}d}$ containing the $\bar{c}$--quark. This sum includes
hadronic states with a large range of invariant mass $M_X$: $M_D \le M_X \le
M_B$. For the bottom mesons, the $b$--quark mass $m_b$ provides a
short--distance scale that leads to a large energy release into the
intermediate hadronic states. Therefore
the energy which flows into hadronic system $X_{c\bar{d}}$ is typically
much larger than the energy scale $\Lambda_{QCD}$ which characterizes the strong
interactions. It will be valid over almost all of the Dalitz
plot, failing only in the narrow corner region where the observed
mass spectrum is dominated by the two narrow $D$ and $D^*$ peaks. Accordingly,
the total SL rate of the $\bar{B}^0$ meson is represented in the
following form
\be
\label{2.1}
\Gamma(\bar{B}^0 \to X_{c\bar d}\ell\nu_{\ell})=
\Gamma(\bar{B}^0 \to D\ell\nu_{\ell})+\Gamma(\bar{B}^0 \to D^*\ell\nu_{\ell})+
\Gamma(\bar{B}^0 \to X'_{c\bar d}\ell\nu_{\ell}),
\ee
where $X'_{\bar{c}d}$ represents the charmed hadron continuum, including also
the resonance states higher than $D$ and $D^*$. The usefulness of such an
expansion rests on large {\it energy release} in the inclusive decay.

Contrary to refs. \cite{MTM96}, \cite{GNST97}, where
the {\it exclusive} decay rates in Eq. (\ref{2.1}) were calculated
using the universal Isgur--Wise (IW) function, our calculations of the exclusive
rates are not relied on the Heavy Quark Effective Theory (HQET).
Instead, they use the
hadronic form factors
that depend on dynamics of
specific channels. The relevant formulae are collected in the Appendix,
see Eqs.(\ref{a.5}), (\ref{a.6}).

Before calculating the {\it inclusive} SL decay rates
we briefly recall the necessary kinematics.
The modulus square of the amplitude summed over the final hadronic states
is written as
\be
\label{2.3}
|M|^2=\frac{G^2_F}{2}|V_{Q'Q}|^2L^{\alpha\beta}W_{\alpha\beta},
\ee
where $V_{Q'Q}$ is the relevant CKM matrix element,
$L^{\alpha\beta}$ is the leptonic tensor
\be
\label{2.4}
L^{\alpha\beta}=8[p_{\ell\alpha}p_{\nu_{\ell}\beta}+p_{\nu_{\ell}\beta}
p_{\ell\alpha}-
g_{\alpha\beta}(p_{\ell}p_{\nu_{\ell}})+i\epsilon^{\alpha\beta\gamma
\delta}p_{\ell}^{\gamma}p_{\nu_{\ell}}^{\delta}],
\ee
with $\epsilon^{0123}=1$, and $W_{\alpha\beta}$ is the hadronic tensor,
\ba
\label{w}
W_{\alpha\beta}&=&(2\pi)^3 \nonumber\\
&&\sum\limits_n\int\limits^{ }_{
}\delta(P_{\bar{B}^0}-q-\sum
\limits^n_{i=1}p_i)<\bar B^0|j_{\alpha}^+(0)|n><n|j_{\beta}(0)|\bar B^0>
\prod\limits^
n_{i=1}\frac{d^3p_i}{(2\pi)^3 2E_i},
\ea
Here, $P_{\bar B^0}$ is the 4--momentum of $\bar B^0$ and 
$q=p_e+p_{\nu}$ is the 4--momentum of the lepton pair.
The hadronic tensor can be decomposed into five different Lorentz
covariants:
\be
\label{2.6}
W_{\alpha\beta}=(-g_{\alpha\beta})W_1+v_{\alpha}v_{\beta}W_2+i
\epsilon_{\alpha\beta\gamma\delta}u^{\gamma}v^{\delta}W_3+
(v_{\alpha}u_{\beta}+u_{\alpha}v_{\beta})W_4+u_{\alpha}u_{\beta}W_5,
\ee
where $v$ is the 4--velocity of decaying $H_Q$, and $u=q/M_{H_Q}$.
The functions $W_i$ depend on two invariants, $q^2$ and
$q_0$, where the latter variable is related to $M_X$ by:
$q_0=(M_{H_Q}^2+q^2-M^2_X)/2M_{H_Q}$.
It is convenient to scale all momenta by $M_{H_Q}$, so that
$q^2=M^2_{H_Q}\cdot t$, $M^2_X=M^2_{H_Q}\cdot s$.
Then the SL width can be cast into the form
\be
\label{2.7}
\Gamma_{SL}=\frac{32}{3}\Gamma_0|V_{Q'Q}|^2\int\limits^{t_{max}}_{t_{min}}
dt\Phi
(t,m^2_1,m^2_2)\int\limits^{s_{max}}_{s_{min}}ds
\frac{|{\bf q}|}{M_{H_Q}}G(t,s),
\ee
where the prefactor
\be
\label{prefact}
\Gamma_0=\frac{G^2_FM^5_{H_Q}}{(4\pi)^3}
\ee
sets the overal scale
of the rate, $\Phi (q^2,m^2_1,m^2_2)$ is the usual triangle function scaled
by $q^2$: $\Phi(q^2,m^2_1,m^2_2)=
\sqrt{1-2\lambda_++\lambda^2_-}$, with $\lambda_{\pm}=(m^2_1\pm m^2_2)/q^2$,
$m_{1,2}$ being the lepton masses, and
\ba
\label{2.8}
G(t,s) &=& 3t(1+\lambda_1-2\lambda_2)W_1(t,s)+\left((1+\lambda_1)
\frac{{\bf q}^2}{M_{H_Q}^2}+\frac{3}{2}\lambda_2t\right)W_2(t,s)+\nonumber\\
&&\frac{3}{2}\lambda_2t\left((1+t-s)W_4(t,s)+tW_5(t,s)\right),
\ea
with $\lambda_1=\lambda_+-2\lambda^2_-$, $\lambda_2=\lambda_+-\lambda^2_-$.
In Eq. (\ref{2.7})
\be
\label{2.9}
\frac{2|{\bf q}|}{M_{H_Q}}\equiv \alpha(t,s)=\sqrt{(1+t-s)^2-4t},
\ee
and the limits of integrations in the $t-s$ plane are given by
\be
\label{2.10}
s_{min}=
\left(\frac{M_X^{(0)}}{M_{H_Q}}\right)^2,~~s_{max}=1-\sqrt{t},~~~~
t_{min}=\frac{(m_1+m_2)^2}{M_{H_Q}^2},~~t_{max}=
\left(1-\frac{M_X^{(0)}}{M_{H_Q}}\right)^2.
\ee
These expressions conclude the kinematical analysis. The next task is the
calculation of the hadronic structure functions $W_i$.

\subsection{The LF constituent quark model approximation for $W_{\mu\nu}$.}

The theoretical treatment of inclusive SL decays of heavy-flavor
mesons carries a distinct similarity to deep--inelastic lepton--nucleon
scattering - this analogy
has been used in refs. \cite{MTM96}, \cite{GNST97} within the framework of
the constituent LF quark model.
The approach is
based on the assumption of quark--hadron duality which means that the sum
over a sufficient number of exclusive hadronic decay modes can be described
in terms of partonic degrees of freedom.
The standard strategy is to represent all the states higher than the ground
pseudoscalar and vector states by the free--quark approximation starting
from some effective continuum threshold $M^{(0)}_X$.
Then the hadronic
tensor $W_{\alpha\beta}$ is given through the optical theorem by the imaginary
part of a quark box diagram describing the forward scattering amplitude
in the second
order in weak interactions
\footnote{
In what follows we consider
the $Q$--
quark
as a virtual particle of the mass $m_Q^2=x^2M_{H_Q}^2$
and neglect the transverse momenta of the $Q$--quark.
As a result,
the expressions for the SL branching ratios
are derived in close analogy with deep--inelastic
scattering. The same result has been
derived \cite{JP95} using the light--cone dominance of SL inclusive $B$--meson
decays, see also \cite{J97}.}:
\be
\label{2.11}
W_{\alpha\beta}=\int^1_0\frac{dx}{x}\int d^2p_{\bot}L^{(Q'Q)}
_{\alpha\beta}(p_{Q'},p_Q)
\delta [(p_Q-q)^2-m_{Q'}^2]\theta(E_{Q'}) |\psi(x, p^2_{\bot})|^2,
\ee
where $E_{Q'}$ is the energy of $Q'$ and the tensor
$L^{(Q'Q)}_{\alpha\beta}$, describing the $Q \to Q'W$ transitions, is
defined analogously to the lepton tensor in Eq.  (\ref{2.4}):  \be
\label{2.12} L^{(Q'Q)}_{\alpha\beta}
=4[p_{Q'\alpha}p_{Q\beta}+p_{Q'\beta}p_{Q\alpha}-
g_{\alpha\beta}(p_{Q'}p_Q)+i\epsilon^{\alpha\beta\gamma
\delta}p_{Q'}^{\gamma}p_Q^{\delta}].
\ee
In Eq. (\ref{2.12}) the extra factor $1/2$ corresponds to the average
over the $b$--quark
spin projections.
Equation (\ref{2.11}) incorporates all long--range QCD effects in the
nonperturbative distribution function $
|\psi(x,p^2_{\bot})|^2$, whose normalization is given by
$\int\limits^1_0dx\int d^2p_{\bot}|\psi(x,p^2_{\bot})|^2=1$, while hard
gluon corrections can be taken into account by perturbative methods and
renormalization--group techniques.
In Eq. (\ref{2.11}) the $\delta$--function corresponds to the decay of a
$Q$--quark with longitudinal momentum $xP_{H_Q}$ to a $Q'$--quark and has two
roots
in $x$, viz.
\be
\label{2.13}
\delta[(p_Q-q)^2-m^2_{Q'}]=\frac{\delta(x-x_+)+\delta(x-x_-)}
{M_{H_Q}^2|x_+-x_-|},
\ee
where
\be
\label{2.14}
x_{\pm}=\frac{1}{2}(1+t-s)\pm\sqrt{(1+t-s)^2-4t
+4\frac{m^2_{Q'}}{M^2_{H_Q}}}.
\ee
By the quark masses $m_Q$ and $m_{Q'}$ we hereafter understand
the "constituent" quark masses taken from a particular constituent quark model.
The root $x_-$ is related to the contribution of
the $Z$--graph arising from the negative energy components of the $Q'$--quark
propagator and is prohibited by
the $\theta(E_{Q'})$ in Eq. (\ref{2.11}). We now substitute into the
quark tensor $L^{(Q'Q)}_{\alpha\beta}$, Eq. (\ref{2.12}),
$p_Q=xP_{H_Q}$, $p_{Q'}=p_Q-q$ and use Eqs. (\ref{2.11}) and
(\ref{2.13}) to obtain
\be
\label{2.15}
W_1=F(x_+),~~
W_2=4\frac{x_+F(x_+)}{|x_+-x_-|},~~
W_3=W_4=-2\frac{F(x_+)}{|x_+-x_-|},~~
W_5=0,
\ee
where
\be
\label{2.16}
F(x)=\int |\psi(x,p^2_{\bot})|^2d^2p_{\bot}.
\ee

After substituting $W_i$ from Eqs. (\ref{2.15}) into  (\ref{2.8})  the
inclusive SL decay rate, Eq. (\ref{2.7}), is given by
\ba
\label{2.17}
\Gamma_{H_{Q'}\ell\nu}&=&\frac{2}{3}\Gamma_0J_{SL}|V_{QQ'}|^2
\int\limits^{t_{max}}_{t_{min}}
dt\Phi(t)\int\limits^{s_{max}}_{s_{min}}\alpha(t,s)\nonumber\\
&&[(1+\lambda_1)\alpha^2(t,s)
\frac{x_+}{x_+-x_-}+3t(1-\lambda_-^2)]F(x_+),
\ea
where we have inserted the factor $J_{SL} \approx 0.9$ representing the effect
of the radiative corrections \cite{NIR89}.
\subsection{Nonleptonic decays.}
The calculation of the NL decay rate closely follows the SL
one. We expect $H_Q$ decays to multimeson states to proceed predominantely
via the formation of a quark--antiquark state, followed by the creation of
the additional $q\bar q$ pairs from the vacuum.
The effective weak Lagrangian, e.g.
for $\bar{b} \to \bar{c} u\bar{q}$ processes with
$q=d,s$ is given by
\be
\label{2.18}
L(\mu)=\frac{G_F}{\sqrt{2}}V_{cb}V_{uq}(c_1O_1+c_2O_2),
\ee
where $O_1$ and $O_2$ denote current--current operators with
the color--nonsinglet and color--singlet structure, respectively:
\be
\label{2.19}
O_1=(\bar{c}\Gamma_{\mu}b)(\bar{q}\Gamma_{\mu}u),~~~
O_2=(\bar{c}_i\Gamma_{\mu}b^j)(\bar{q}_j\Gamma_{\mu}u^i),
\ee
and  $\Gamma_{\mu}=(1-\gamma_5)\gamma_{\mu}.$
The lepton pair is substituted by a quark pair, and the Wilson
coeffients $c_i$ are the perturbative QCD corrections
that describe the physics between the $W$
boson mass and the characteristic hadronic scale of the process.
We shall use the values \cite{BB93}
\be
\label{2.21}
c_1(m_b)=1.132,~~c_2(m_b)=-0.286;~~~~c_1(m_c)=1.351,~~c_2(m_c)=-0.631,
\ee
obtained at next--to--leading order with the evolution
of the running coupling constant being done at two--loop level using the
normalization $\alpha_s(m_Z)=0.118 \pm 0.003$. The {\it inclusive} NL rate
is given by Eq. (\ref{2.17}) with the substitution $\Gamma_0J_{SL} \to
3\Gamma_0\eta $, where
\be
\label{2.22}
\eta=c^2_+\frac{N_c+1}{2N_c}+c^2_-\frac{N_c-1}{2N_c},
\ee
with $c_{\pm}=c_1 \pm c_2$. Following ref. \cite{GNST97} we let $N_c \to
\infty$ in Eq. (\ref{2.22}) in which case $\eta$ is reduced to
\be
\label{2.23}
\eta=\frac{1}{2}(c^2_++c^2_-).
\ee
For the exclusive two--meson NL decays $H_Q\to PP,~PV,~VV$, where $P$ and
$V$ are the lowest--lying pseudoscalar and vector mesons, respectively,
we use the BSW approach \cite{BSW85}, \cite{BSW87}. There are two main ingredients
in this approach:\\
\begin{itemize}
\item one assumes factorization:
$<M_1M_2|J_{\mu}J_{\mu}|H_Q>
 =<M_2|J_{\mu}|0><M_1|J_{\mu}|H_Q>$ to describe
$H_Q \to M_1M_2$ decay,
\item one employs  LF meson wave functions to compute
$<M_1|J_{\mu}|H_Q>$.
\end{itemize}
This method
which lacks a firm footing for the $B$  meson, can in fact be
justified in the case of $B_c$. Deviations from factorization
arise from higher Fock components of the $B_c$ wave function
and therefore are of higher order in the nonrelativistic
expansion.

By factorizing matrix elements of 4--quark operators contained in
(\ref{2.18}), we distinguish two classes of
NL decays \cite{BSW87} corresponding to two flavor--flow topologies
relevant for our discussion: the so--called "tree topology" (class I) and
the "color--suppressed tree topology" (class II).
The class I ({\it external} decays) contains those decays
where only a {\it charged} meson can be generated directly from a
colour--singlet current.
As before, we calculate separately the rates for the exclusive two--meson
decays of the type
$\bar{B}^0 \to D^+\pi^-$. $\bar{B}^0 \to D^+\pi^-$,
$\bar{B}^0 \to D^+\rho^-$, $\bar{B}^0 \to D^+\rho^-$, etc., and the rates
for the decays into multimeson states like
$\bar{B}^0 \to X'_{\bar{d}c}\pi^-$, $\bar{B}^0 \to D^+X'_{\bar{u}d}$, or
$\bar{B}^0 \to X'_{\bar{d}c}+X'_{\bar{u}d}$.
A second class of transitions (internal decays) consists of those decays
in which the {\it neutral} meson is generated from the quark current,
like $J/\psi$ in the decay $\bar{B}^0 \to \bar{K}J/\psi$.
There is the third class of
decays where the external and internal amplitudes interfere. These decays
play important role in the case of the $D$ mesons, but are less important
for $B_c$, and are absent for $\bar{B}^0$. The class I and class II
two--meson amplitudes are proportional to the QCD coefficients
\be
\label{a1}
a_1=c_1(\mu)+\frac{c_2(\mu)}{N_c},~~a_2=c_2(\mu)+\frac{c_1(\mu)}{N_c},
\ee
repectively, with $ \mu=O(m_b)$.
The term proportional to $1/N_c$ arises from the Fierz reordering
of operators $Q_2$ to produce quark currents to match the quark
content of the hadrons in the initial and final state after adopting the
factorization assumption. This well--known procedure \cite{BSW87} results in
matrix elements with the right flavor quantum number but involve both
color--singlet and color--octet quantum operators. We shall use the "naive"
factorization approximation in which one discards the color--octet operators
\footnote{To compensate for the neglected octet operators and other non--
factorizing contributions one usually treats $\xi=1/N_c$ in Eq. (\ref{a1})
as a free parameter which can, in principle, be channel dependent. For a detailed
discussion see refs. \cite{S97}}. In what follows we neglect the $1/N_c$
corrections in Eqs. (\ref{a1}) letting $a_1=c_1$ and $a_2=c_2$.

In conclusion, we list the expressions used below to   calculate
 the NL partial widths. The {\it exclusive} two--meson decay rates
are calculated
using Eqs. (\ref{a.15})--(\ref{a.18}) given in Appendix. The {\it exclusive} NL rates for the decays
of the type $\bar{B}^0 \to D^+X'_{\bar{u}d}$ are given by Eqs.
(\ref{a.5}),(\ref{a.6}). The {\it inclusive} NL rate of the type
$\bar{B}^0 \to \pi^-(\rho^-)X_{\bar{d}c}$ are given by Eqs.
(\ref{a.25}),(\ref{a.26}).
Finally, the {\it inclusive}
NL rates, like $\bar{B}^0 \to X'_{\bar{d}c}+X'_{\bar{u}d}$,  have been
 calculated using Eq. (\ref{2.17}), where the lepton pair is
substituted by a quark pair and the prefactor $\Gamma_0$ is replaced
by $3\Gamma_0c^2_1$
or $3\Gamma_0c^2_2$ for the external and internal  inclusive
decays, respectively.
\section{NUMERICAL RESULTS}
We are now ready to add the various contributions presented above
and to estimate the $B_c$ lifetime.
We first summarize the input parameters that will be used in
our numerical calculations. The partial decay rates depend on the following
set of parameters:
\begin{itemize}
\item the masses of constituent quarks building up final mesons and
multi--
hadrons and the masses of pseudoscalar and vector mesons; the formers  are
taken from the ISGW2 model \cite{IS95}:
\be
m_u=m_d=0.33,~~m_s=0.55,~~m_c=1.82,~~m_b=5.2,
\ee
while the latters have been taken from the Particle Data Group
publication \cite{PDG96}.
Note that the {\it constituent} quark masses $m_b$ and $m_c$ satisfy
approximately the relation $m_b=m_c+3.4~GeV$, which is consistent with
the known formula relating the {\it pole} masses $m_{b,pole}$ and
$m_{c,pole}$ in HQET;
\item the meson decay constants $f_{PS}$ and $f_V$ of the pseudoscalar and vector mesons
to the $W$--boson. For a pseudoscalar meson $P=(q_1\bar{q}_2)$, we define
$<0|\bar{q}_2\gamma_{\mu}\gamma_5q_1|P(p)>=if_{PS}p_{\mu}$.
The decay
constant of a vector meson $V=(q_1\bar{q}_2)$ is defined as
$<0|\bar{q}_2\gamma_{\mu}\gamma_5q_1|V>=\varepsilon_{\mu}m_Vf_V$.
The constants $f_{PS}$ and $f_V$ used in our calculations were taken from ref.
\cite{NS97}. The $B_c$ meson mass is known
with a good accuracy from the current quark model calculations
calculations, we use $M_{B_c}=6.3 GeV$;
\item the threshold values $M^{(0)}_X$ at which
the hadron continuum starts; in our calculations we have adopted two different
choices:
$M_X^{(0)}=M_{P}+m_{\pi}$ (case $a$)
and
$M_X^{(0)}=M_V+m_{\pi}$ (case $b$),
where $M_{P}$ and $M_V$ are the masses of the
pseudoscalar and vector ground state mesons, respectively.
\end{itemize}

\noindent For the radial wave function appearing in Eq. (\ref{1.2}), the Gaussian
ans\"atz of the ISGW2 model has been adopted in which the main characteristic
is confinement. The oscillator parameters have been taken from \cite{IS95}.
Their values along with the values of the meson decay constants are collected
in Table 1.
\subsection{$\bar{B}^0$ decays}
First, we would like to present an overview of different $B_c$--decay mechanisms
and their relative importance as obtained within the framework we are
advocating. To this end, we first calculate the partial $\bar{B}_0$ decay
modes corresponding to various underlying quark subprocesses.
In Table 2 we show the $\bar{B}^0$ partial widths for
two specific choices of the parameter $M^{(0)}_X$. The left set of numbers
collects numerical results obtained for the threshold value for the hadron
continuum $M^{(0)}_X=M_P+m_{\pi}$, while the right set is given for
$M_X^{(0)}=M_V+m_{\pi}$, where in the considered case $M_P=M_D$ and
$M_V=M_{D^*}$ \footnote{Note that for the former choice there is, in
principle, a danger of double counting, because e.g. the exclusive rate
${B}^0 \to D^0$ accounts partially for the two--particle
state $D\pi$, which in our approach is included in the inclusive rate}.
Table 2 also shows the numbers $N_{exc}$ and $N_{imc}$ of (both {\it external}
and {\it internal}) exclusive and
inclusive channels, respectively, for the $\bar b\to \bar c$ decays.
Our analysis
incorporates 54 exclusive SL and NL
$\bar{b} \to \bar{c}$ decays and 29 inclusive $\bar{b} \to \bar{c}$
decays including two baryon-antibaryon channels. The latters
were calculated using the Stech approach \cite{S87}. We have also included the
CKM suppressed $\bar b\to \bar u$ contributions (with $|V_{ub}/V_{cb}|=0.08$)
which slightly change the overall
results. We take the vector and axial form factors for $\bar{B}^0 \to
D(D^*)\ell\nu_{\ell}$ and $\bar{B}^0 \to \pi(\rho)\ell\nu_{\ell}$ transitions
from ref.  \cite{DKND97}, where the fit of the form factors by the functions
\be f(q^2)=\frac{f(0)}{1-q^2/\Lambda^2_1-\beta \cdot q^4/\Lambda^4_2}, \ee
where $f=F_1,~V,~A_0,~A_1,~A_2$
have been also presented. The distribution functions
$F_b(x)$ for the $\bar{B}^0$ meson and $F_{\bar{b}}(x)$ for the $B_c$ meson
are shown in Fig. 1.  Note that in the later case $F_c(x)=F_{\bar{b}}(1-x)$.
In terms of the mean
value $<x>=\int\limits^1_0dx~x~F(x)$ and the variance $\sigma^2=\int\limits^1_0
dx~(x-<x>)^2~F(x)$, one gets
$<x>=0.90$, $\sqrt{\sigma^2}=0.06$ for $\bar{B}^0$ and
$<x>=0.72$, $\sqrt{\sigma^2}=0.09$ for $B_c$. The difference in the values of
$\sigma^2$  reflects the difference in the values of $\sqrt{<p_{\bot}^2>}=
\beta_{H_Q}$
for $\bar{B}^0$ and $B_c$.

 Numerical values of NL branchings depend upon estimation
 of QCD corrections which give contribution of order of $10\%-20\%$\footnote
{The treatment of radiative corrections in our phenomenological application
 is associated with
large uncertainties. In a typical hadronic process, there are several mass
scales involved: the hadron masses, the quark masses, the energy release,
etc. Thus, there is an uncertainty in the choice of the ``characteristic''
scale $\mu$ of a process. In principle this is not a problem, since the products of the Wilson coefficients with the hadronic matrix elements are scale
independent. However, we employ simple model estimations of the matrix elements, which do not yield an explicit scale dependence that could compensate that of the Wilson coefficients. Instead, the quark model calculations are assumed
to be valid on a particular scale, typically $\mu \approx 1-2~GeV$.}.
 The main uncertainty ($\approx 10 \%$) is related to the choice of
$M_X^{(0)}$. For each case we then fix the effective value of $|V_{cb}| $
by the requirement that the measured $\bar B^0$ meson lifetime
$\tau_{\bar B^0} \approx 1.56~ps $ is obtained.
Our aim here is not to
establish a new value of $|V_{cb}|$, but rather to illustrate how our
approach works. Moreover, imposing
the $|V_{cb}|$ constraint strongly reduces the dependence of the
predicted value of $\tau_{B_c}$ from the uncertainty
related to the choice
of the continuum threshold
\footnote{In addition to using the measured $\bar{B}^0$
lifetime one could be tempted to use the $D^0$ lifetime to eliminate the
similar uncertainties for the $c\to s$ transitions. However, we refrain from
this procedure, since our approach applied to charm mesons is less reliable
than for $B_c$ and probably more qualitative than quantitative.}.

A few tests of the model form factors may be quoted here: The branching
ratios for the SL decays
$\bar{B}^0\to D^*\ell\bar{\nu}$ and $\bar{B}^0\to D\ell\bar{\nu}$ are found
to be $5.0\%$ and $1.9\%$, respectively. The corresponding
experimental values are $(4.64\pm0.26)\%$ and $(1.8\pm0.4)\%$
\cite{PDG96}. In the SL $\bar{B^0} \to \rho\ell\bar{\nu}$, the values
for the form factors at $q^2=0$ are $V=0.216$, $A_1=0.170$,
$A_2=0.155$.
The expected branching ratio for this decay is
$13.2|V_{ub}|^2$, which when combined with experimental result
$BR(\bar{B}^0 \to \rho\ell\bar{\nu_{\ell}}= (2.5^{+0.8}_{-09})
\cdot 10^{-4}$ \cite{JPA96} yields $|V_{ub}|=(3.1\pm0.5)\cdot 10^{-3}$.
The calculated
$\rho/\pi$ ratio in exclusive SL $\bar{B}$ decays is $2.2$, as
compared with the experimental value of $1.4\pm0.6$.

Our predictions for the branching ratios of the dominant NL
two--body decays of the $\bar{B}^0$ meson are given in Table 3. The QCD
coefficients $a_1$ and $a_2$ have been left as parameters in the
expressions for the branching ratios. For comparison, we show the corresponding
predictions obtained using the Neubert--Stech \cite{NS97} and NRSX models
\cite{NRSX92} and the world average experimental results, as recently compiled
in \cite{BHP96}.

It is instructive to compare in details some theoretical predictions
for the NL two--body decays with
the data. Due to the uncertainty in the values of the QCD parameters $a_1$ and
$a_2$, we first concentrate on the ratios of branching fractions in which these
coefficients cancel. From the class I transitions listed in Table 3 we obtain
\be
R_1=\frac{BR(\bar{B}^0 \to D^+\pi^-)}{BR(\bar{B}^0 \to D^{*+}\pi^-)}
\approx 1.06~
[1.04,~1.07],
\ee
and
\be
R_2=\frac{BR(\bar{B}^0 \to D^+\rho^-)}{BR(\bar{B}^0 \to D^{*+}\rho^-)}
\approx 0.94~
[0.88,~0.89].
\ee
Here we use the LF ISGW2 model as our nominal choice, and quote
results obtained with the models of refs. \cite{NS97}, \cite{NRSX92} in
brackets.  The experimental results for these quantities $R_1=1.11\pm0.23$ and
$R_2= 1.15\pm0.34$ agree with the predictions and do not distinguish between
the models because the errors are still
large. For the class I transitions to final states which differ only in their
light mesons we obtain
\be
R_3=\frac{f^2_{\rho}}{f^2_{\pi}}\cdot
\frac{BR(\bar{B}^0 \to D^+\pi^-)}{BR(
\bar{B}^0 \to D^{*+}\rho^-)}\approx 1.03~
[1.03,~1.06],
\ee
and
\be
R_4=\frac{f^2_{\rho}}{f^2_{\pi}}\cdot
\frac{BR(\bar{B}^0 \to D^{*+}\pi^-)}{BR(
\bar{B}^0 \to D^{*+}\rho^-)}\approx 0.91~
[0.88,~0.88].
\ee
The experimental values are
$R_3=0.95\pm0.24$ and $R_4=0.99\pm0.25$\footnote{Note that in the
heavy--quark limit the fraction ratios $R_i$, $i=1,...4$ equal unity.}, again
in agreement with our predictions.
Similar ratios can be taken for class II amplitudes. Based on the results
of our model, we expect
\be
R_5=\frac{BR(\bar{B}^0 \to \bar{K}J/\psi)}
{BR(\bar{B}^0 \to \bar{K}^*J/\psi)}
\approx 0.44~[0.62,~0.32].
\ee
The corresponding experimental value
is $R_5=0.58\pm0.11$. \\

\subsection{$B_c$ decays}

The transition operators driving $B_c$ decays are the same that generate
$B$ and $D$ decays. However their expectation values are evaluated for the
$B_c$ wave function, rather than the $B$ and $D$ wave functions reflecting
that $\bar b\to \bar c$ and $c \to s$ transitions proceed in a
different environment.
This is illustrated by the results of Table 4, where we compare
various $B_c$ form factors at $q^2=0$ with the corresponding
$\bar{B}^0$ and $D^0$ form factors.

We apply the strategy outlined in
the previous sections to calculate different partial rates and the
lifetime of the $B_c$.  The critical point with regards to this issue
is charm decay.  Here the energy release is not as comfortably large
as it is in the case of bottom decay. As a result, our estimations of
the inclusive charm decay should be more sensitive to a hadronization
model.  However, the inclusive charm decay contributes only $8 \%$ to the total
$c\to s$ rate of $B_c$.  For this reason we do not include any hadronization
corrections in our calculations.

The results for the partial $B_c$ decay modes corresponding to the various
underlying quark subprocesses are collected in Table 5 for $M^{(0)}_X=M_{P}+
m_{\pi}$. As in Table 2 we present the numbers of the exclusive and
inclusive channels included in the calculations. The number of the {\it
exclusive} $\bar{b} \to \bar{c}$ decays in the $B_c$ case is
less than that for $\bar{B}^0$ because of the interference effects in the
class $III$ amplitudes (see Table
7)\footnote{The sum of the other interference effects is effectively accounted
for by the Pauli interference contribution $\Gamma_{PI}$, see below}.
For comparison we also show
in Table 5 the results obtained in \cite{BB96}, using the
nonrelativistic QCD. Viewing this comparison with due caution,
regarding the model dependence and other uncertainties in the
estimation of the decay modes as well as the quark mass uncertainty
for the inclusive prediction, it is reassuring that the order of
magnitude comes out to be consistent. Our bound state corrections are
numerically a bit larger than very small effects reported in
\cite{BB96}.  For the sum of  $\bar{b} \to \bar{c}$
 spectator contributions we obtain
$\Gamma^{(\bar{b} \to \bar{c})}=0.83\cdot \Gamma(\bar{B}^0)=
0.531~ps^{-1}$,
while for the total $c$--decay contribution one finds
$\Gamma^{(c \to s)} = 0.38\cdot \Gamma(D^0)= 0.915 ~ps^{-1}$. For the choice
$M_X^{(0)}=M_V+m_{\pi}$ we obtain $\Gamma^{(\bar{b}\to \bar{c})}=0.79\cdot \Gamma(\bar{B}^0)=
0.504ps^{-1}$ and $\Gamma^{(c\to s)}=0.43\cdot \Gamma(D^0)=1.049~ps^{-1}$
\footnote{
The $\bar{b}\to \bar{c}$ rates have been calculated usig  the values
of the effective CKM parameter $|V_{cb}|$ listed in Table 2 for
cases $a$ and $b$, respectively}.
Various branching fractions can be also inferred from
Table 5. For instance the SL branching ratio $BR(B_c \to e\nu X)$ is
found to be $9.3\%$, i.e.  $\approx 15 \%$ less than $BR(\bar B^0 \to
e\nu X)$.  Details of our predictions for the partial widths are
presented in Table 6. In particular, we obtain \be
\Gamma(B_c\to e\nu_eX_{\bar{c}c})=4.1\cdot 10^{13}|V_{cb}|^2sec^{-1},
\ee
\be
\Gamma(B_c\to e\nu_eX_{\bar{b}s})=8.5\cdot 10^{10}|V_{cs}|^2sec^{-1}.
\ee
The corresponding results of the original ISGW2 model  are \cite{IS95}
\be
\Gamma(B_c\to e\nu_eX_{\bar{c}c})=3.36\cdot 10^{13}|V_{cb}|^2sec^{-1},
\ee
\be
\Gamma(B_c\to e\nu_eX_{\bar{b}s})=5.0\cdot 10^{10}|V_{cs}|^2sec^{-1}.
\ee
The ratio of $B_c$ decay {\it via} $\bar b\to \bar c$ to $c\to s$ decay is
\be
\frac{\Gamma(B_c\to e\nu_eX_{\bar{b}s})}{\Gamma(B_c\to
e\nu_eX_{\bar{c}c})}=0.0021\frac{|V_{cs}|^2}{|V_{bc}|^2}\approx 1.3,
\ee
The SL
width
$\Gamma(B_c \to e\nu_eX_{\bar{c}c})=0.061~ps^{-1}$
is only $\approx 20\%$ smaller than  that of $\bar{B}^0$,
$\Gamma(\bar B^0 \to e\nu_eX_{\bar{c}d})=0.074~ps^{-1}$,
even though the spectator $c$ quark is no longer
light. The contributions from the pseudoscalar and vector final states
are $21\%$ and $44\%$, that is not too much different from the predictions
 of the ISGW2 model.
For the SL $c\to s$ decays the recoil effects are very small due to the
large daughter mass. The SL width
$\Gamma(B_c \to e\nu_eX_{\bar{b}s})=0.081~ps^{-1}$ is dominated by the decays to
the $B^0_s$ $(27.2\%)$ and $B^{0*}_s$ $(63.2\%)$, because the avaiable energy
is small.

In our phenomenological description we also include non--spectator contributions
from weak annihilation (WA) and Pauli interference (PI) \cite{VS87}. The
the contribution of the annihilation channel is
\be
\label{ann}
\Gamma_{a}=\sum_{i=\tau,c}\frac{G^2_F}{8\pi}|V_{cb}|^2M_{B_c}^5
\left(\frac{f_{B_c}}{M_{B_c}}\right)^2\left(\frac{m_i}{M_{B_c}}\right)^2
\left (1-(\frac{m_i^2}{M_{B_c}})^2\right)^2\cdot \tilde{c}_i,
\ee
where $f_{B_c}=0.42~$Gev,~$\tilde{c}_{\tau}=1$ for the $\tau^+\nu_{\tau}$
channel and $\tilde{c}=(2c_+(2\mu_{red})+c_-(2\mu_{red}))^2/3$ for
the $c\bar s$ channel, with $\mu_{red}=m_bm_c/(m_b+m_c)$ being the
reduced mass of the $\bar{b}c$ system. The result is known to two--loop
accuracy \cite{BW90}: $c_+(2\mu_{red})=0.8$,
$c_-(2\mu_{red})=1.5$. The leptonic
constant $f_{B_c}$ has been estimated through the LF wave function
$\psi(x,p^2_{\bot})$ using Eq. (13) of ref. \cite{roma94}.
Note that because of the partial cancellation of weak
annihilation rate and the effect of the Pauli interference diagrams,
no significant uncertainty on the lifetime arises from the limited
knowledge of the decay constant $f_{B_c}$. We find
$\Gamma_a=0.189~ps^{-1}$ and $\Gamma_{PI}= -0.100~ps^{-1}$. The similar
value of $\Gamma_{PI}$ has been obtained in \cite{BB96}.
Putting everything
together we obtain for the $B_c$ width $\Gamma(B_c)=
\Gamma^{\bar{b}\to \bar{c}}(B_c)+
\Gamma^{c\to s}(B_c)+\Gamma_a+\Gamma_{PI}$:
\be
\label{Gamma}
\Gamma(B_c)=(0.65~ps)^{-1},~case~a;~~~
\Gamma(B_c)=(0.61~ps)^{-1},~case~b.
\ee
One observes a dominance of the charm decay modes over $b$--quark decays.
We consider
the dispersion in predicted values of $\tau_{B_c}$ as a rough measure of our theoretical
uncertainty in calculation of the inclusive decay rates, therefore our final result is
\be
\label{FR}
\tau_{B_c}=0.63 \pm 0.02 ~ ps^{-1}.
\ee

Note that the {\it exclusive} $\bar b$-- and $c$--decay rates
are $0.32~ps^{-1}$ and $0.80~ps^{-1}$, i.e. $52\%$ and $87\%$
of the total $\bar b\to \bar c$ and $c\to s$ decay rates, respectively, In
Tables 7,8 we compare our results for the exclusive two--body NL
$B_c$ decay modes
with the results obtained using the BSW and ISGW models \cite{LM91} and with
the results of ref. \cite
{CC94}. Adding the rates for twenty of these two--body decay modes one
finds $0.69~ps^{-1}$ (BSW model) and $0.90~ps^{-1}$ (ISGW model) \cite{LM91}
and $1.15~ps^{-1}$ in \cite{CC94}, to be compared with our result
$0.70~ps^{-1}$ which is the sum of the rates of $0.67~ps^{-1}$ for
$c\to s$ transitions and $0.03~ps^{-1}$ for $b\to c$ transitions
from Table 7,8.

Finally, we note that the experimental extraction of $B_c$ signal from the
hadronic background
requires the reliable estimation of the branching fraction $
BR(B_c\to J/\psi+X) $, because $J/\psi$ can be easily identified by
its leptonic decay mode, while the experimental registration of the
final states containing the $\eta_c$ or $B^{(*)}_s$ is impeded by the
large hadron background.  We obtain
$BR(B_c\to J/\psi+\mu\nu_{\mu})=1.7\%$,
$BR(B_c\to J/\psi+\pi)=0.1\%$, and $BR(B_c\to J/\psi+X)=13.2\%$.

\section{Summary}

In conclusion, we have used a relativistic constituent quark model based
on the LF formalism to perform a detailed investigation of the $B_c$ meson
partial widths and the lifetime of the $B_c$ meson. The hadronic form factors
and the distribution function were calculated using meson wave functions
derived from an effective $q\bar q$ interaction intended originally to
describe the meson mass spectra. In this way the link between $B_c$ physics
and the "spectroscopic" constituent quark models was explicitly established.
For numerical estimates we employed the LF quark functions that are related
to the equal--time wave functions of the ISGW2 model.
In several important
aspects our analysis goes beyond the quark model estimations derived
previously in the literature. In addition to the frequently discussed
SL and the
two--meson {\it exclusive} decay modes, we
have included 84 exclusive decay modes and 44 SL and NL {\it inclusive}
decay modes, that increases the total $B_c$ width by $\approx 30 \%$.  Account
of the inclusive decays leads to a considerably smaller $B_c$--lifetime,
$\tau_{B_c}=0.63\pm0.02~ps$, which is in fairly good agreement with the estimate
obtained non--relativistic
QCD, $\tau_{B_c}=(0.55 \pm 0.15)~ps$ \cite{BB96} and also agrees with the most recent
CDF measurement \cite{CDF98} within one standard deviation.

To sum up, the LF constituent quark model makes clear predictions on the
global pattern: (i) a short $B_c$ lifetime well below $1~ps$ and (ii) a
predominance of charm over beauty decays among the NL modes.

\setcounter{equation}{0}
\def\theequation{A.\arabic{equation}}

\section*{Appendix}
The vector and axial hadronic form factors relevant for weak SL
decays of a
pseudoscalar meson $H_Q$, containing a heavy quark $Q$, to a member of
the lowest--lying multiplet of pseudoscalar or vector mesons are defined
in a usual way (see e.g. \cite{NS97}).
We denote by $P_{H_Q}$, $P_{H_{Q'}}$ and $M_{H_Q}$, $M_{H_{Q'}}$
the 4--momenta and
masses of the parent
and daughter meson, respectively. Then the amplitude
$<P_{H_{Q'}}|V_{\alpha}|P_{H_Q}>$
can be expressed in terms of two form factors

\be
\label{a.1}
<P_{H_{Q'}}|V_{\alpha}|P_{H_{Q}}>=(P_{\alpha}-
\frac{M_{H_Q}^2-M_{H_{Q'}}^2}{q^2}q_{\alpha})F_1(q^2)+
\frac{M_{H_Q}^2-M_{H_{Q'}}^2}{q^2}
q_{\alpha}F_0(q^2),
\ee
where $P=P_{H_Q}+P_{H_{Q'}}$, $q=P_{H_Q}-P_{H_{Q'}}$.
There is one form factor for the amplitude\\
$<P_{H_{Q'}},\varepsilon|V_{\alpha}|P_{H_Q}>$
\be
\label{a.2}
<P_{H_{Q'}},\varepsilon|V_{\alpha}|P_{H_Q}>=
\frac{2i}{M_{H_Q}(1+\zeta)}\epsilon_{\alpha\beta\gamma\delta}\varepsilon^{*\beta}P_1^
{\gamma}P_2^{\delta}V(q^2),
\ee
where $\zeta=\frac{M_{H_{Q'}}}{M_{H_Q}}$\footnote{
In what follows we suppress the indeces 1 and  2 (corresponding 
to the meson containing a quark--spectator and to the one generated from 
a $W$ current, respectively) of $\zeta$ everywhere except in Eq. (\ref{a.19}).} 
and three independent form factors
for the amplitude $<P_{H_{Q'}},\varepsilon|A_{\alpha}|P_{H_Q}>$

\ba
\label{a.3}
<P_{H_{Q'}},\varepsilon|A_{\alpha}|P_{H_Q}>&=&M_{H_Q}(1+\zeta)\varepsilon^{*\beta}
A_1(q^2)-
\frac{(\varepsilon^*q)}{M_{H_Q}+M_{H_{Q'}}}P_{\alpha}A_2(q^2)-\nonumber\\
&&2M_{H_{Q'}}\frac{(\varepsilon^*q)}{q^2}
q_{\alpha}A_3(q^2)+
2M_{H_Q}\frac{\varepsilon^*q}{q^2}A_0(q^2),
\ea
where $A_0(0)=A_3(0)$ and the form factor $A_3(q^2)$ is given by the linear
combination
\be
\label{a.4}
A_3(q^2)=\frac{1}{2\zeta}\left((1+\zeta)A_1(q^2)-(1-\zeta)A_2(q^2)\right).
\ee
We view the form factors as
functions of the dimensionless variable $y=v_{H_Q}\cdot v_{H_{Q'}}$, where
$P_{H_Q}=M_{H_Q}\cdot v_{H_Q}$,
$P_{H_{Q'}}=M_{H_{Q'}}\cdot v_{H_{Q'}}$,
and $q^2=M_{H_Q}^2+M^2_{H_{Q'}}-2M_{H_Q}
M_{H_{Q'}}y$
\footnote{Although we are using the variable $v_1\cdot v_2$ we are not treating the daughter
quark as heavy.}.
The differential rates for a decay of a pseudoscalar particle into
another pseudoscalar
particle and lepton or quark pair is given by
\ba
\label{a.5}
\frac{d\Gamma}{dy}&=&\frac{16}{3}\Gamma_0c^2|V_{Q'Q}|^2|V_{q_1q_2}|^2
\zeta^4
\Phi(q^2,m_1^2,m_2^2)J\cdot \nonumber\\
&&\cdot \sqrt{y^2-1}[(1-\frac{\lambda_1}{2})
F_1^2(y)(y^2-1)+
\frac{3}{8}\lambda_2(\zeta^{-2}-1)F_0^2(y)]
\ea
where $V_{Q'Q}$ and $V_{q_1q_2}$ are the corresponding CKM matrix elements
and the factor $J$ takes into account  the perturbative QCD corrections.
For the SL decays $c^2=1$, $V_{q_1q_2}=1$, and $J\approx 0.9$;
for the external (internal) NL decays $c^2=3c_1^2~(3c^2_2)$ and
$J=1$.  The differential transition rate to a
vector particle is

\ba
\label{a.6}
\frac{d\Gamma}{dy}&=&4\Gamma_0c^2|V_{Q'Q}|^2|V_{q_1q_2}|^2
\zeta^2
\Phi(q^2,m_1^2,m_2^2)J\cdot \nonumber\\
&&\cdot \sqrt{y^2-1}[(1-\frac{\lambda_1}{2})
H^2(y)+
6\lambda_2\zeta^2(y^2-1)A^2_0(y)],
\ea
where
\be
\label{H}
H^2(y)=\frac{t}{M_{H_Q}^2}(H_+^2(y)+H_-^2(y)+H_0^2(y)),
\ee
and  the helicity amplitudes are given by
\be
\label{a.7}
H_{\pm}(y)=M_{H_Q}\left((1+\zeta)A_1(y)\mp 2\sqrt{y^2 -1}
\frac{\zeta}{1+\zeta}V(y)\right),
\ee
\be
\label{a.8}
H_0(y)=\frac{M_{H_Q}^2}{\sqrt{q^2}}\left((y-\zeta)(1+\zeta)A_1(y)-2(y^2-1)
\frac{\zeta}{1+\zeta}A_2(y)\right)
\ee
The maximal value of $y$ is $(M_{H_Q}^2+M_{H_{Q'}}^2)/(2M_{H_Q}M_{H_{Q'}})$,
corresponding to $q^2=0$.

Recall that in the case of the heavy--to--heavy transitions $H_Q \to H_{Q'}$
the heavy quark symmetry implies simple relations between various form
factors:
\be
F_1(y)=V(y)=A_0(y)=\frac{1+\zeta}{2\sqrt{\zeta}}\xi_{IW}(y),
\ee
\be
F_0(y)=A_1(y)=\frac{2\sqrt{\zeta}}{1+\zeta}\frac{y+1}{2}\xi_{IW}(y),
\ee
where $\xi_{IW}(y)$ is the Isgur--Wise function. For the realistic
case of finite heavy--quark masses, these relations are
modified by corrections that break heavy quark symmetry. These corrections have been
analyzed for $\bar B^0$ and $D$ decays using the LF technique
\cite{GNS96}, \cite{DKND97}.  We use the same technique to calculate
various form factors for the $B_c$ decays. The values of these form
factors at $q^2=0$ have been already given in Table 4.

The NL two--body partial widths are given by expressions
\be
\label{a.15}
\Gamma_{P_2P_1}=\frac{c^2}{2}\cdot\Gamma_0
|V_{Q'Q}|^2
|V_{q_1q_2}|^2
F^2_{PS}
\cdot(1-\zeta^2)^2
\alpha
F_0^2(\zeta^2),
\ee
\be
\label{a.16}
\Gamma_{V_2P_1}=\frac{c^2}{2}\cdot\Gamma_0|V_{Q'Q}|^2
|V_{q_1q_2}|^2
F^2_V\cdot
\alpha^3
F_1^2(\zeta^2),
\ee
\be
\label{a.17}
\Gamma_{P_2V_1}=\frac{c^2}{2}\cdot\Gamma_0|V_{Q'Q}|^2
|V_{q_1q_2}|^2F^2_{PS}\cdot
\alpha^3
A_0^2(\zeta^2),
\ee
\be
\label{a.18}
\Gamma_{V_2V_1}=\frac{c^2}{2}\cdot\Gamma_0|V_{Q'Q}|^2
|V_{q_1q_2}|^2
F^2_V\cdot
\alpha
H^2(\zeta^2),
\ee
where
\be
\label{a.20}
F_{PS}=\frac{2\pi f_{PS}}{M_{H_Q}},~~~~F_V=\frac{2\pi f_V}{M_{H_Q}},
\ee
\be
\label{a.19}
\alpha \equiv
\alpha(\zeta_1^2,\zeta_2^2)=\sqrt{(1+\zeta^2_1-\zeta^2_2)^2-4\zeta^2_1},
\ee
and $c=a_1$ for the {\it external} two--body decays, $c=a_2$ for the {\it
internal} two--body decays.

Finally, the partial widths of {\it inclusive} NL decays in which the final state
contains a charged or neutral meson directly generated by a color--singlet
current are given by

\ba
\label{a.25}
\frac{d\Gamma_{PX}}{ds}&=&\Gamma_0\frac{c^2F^2_P}{4}|V_{Q'Q}|^2|V_{q_1q_2}|^2
\cdot\nonumber\\
&&[(1+s-\zeta^2)^2
W_2-4\zeta^2W_1+4\zeta^2(1+s-\zeta^2)W_4]
\alpha(\zeta^2,s)
\ea
\ba
\label{a.26}
\frac{d\Gamma_{VX}}{ds}&=&\Gamma_0\frac{c^2F^2_V}{4}|V_{Q'Q}|^2|V_{q_1q_2}|^2
\cdot
[\alpha^2(\zeta^2,s)W_2+12\zeta^2W_1]
\alpha(\zeta^2,s)
\ea
\section*{Acknowledgements}
One of the authors (I.M.N.) would like to thank Dr. Y.--Y. Keum and APCTP for
the warm hospitality. This work was supported by the INTAS--RFBR grants,
refs. No 95--1300 and 96--155, and the RFBR grant, ref. No 95--02--04808a.

\vspace{0.5cm}
\noindent {\bf Table 1.} The oscillator parameters $\beta$ for various mesons
\cite{IS95} and the meson
constants $f_{PS}$, $f_V$ \cite{NS97} used in the calculations.\\[3mm]
\vspace{0.2cm}
\begin{tabular}{|c|c|c|c|c|c|c|c|c|c|c|c|c|c|}
\hline\hline
Meson & $\pi$ & $K$ & $\eta_c$ & $D$ & $D_s$ & $\bar{B}^0$ & $B_s$ &
$B_c$ &$\rho$ & $K^*$ & $J/\psi$ & $D^*$ & $D^*_s$ \\ \hline\hline
$\beta$
& 0.41 & 0.44 & 0.88 & 0.45 & 0.56 & 0.43 & 0.54 & 0.92 & 0.30 & 0.33
& 0.62 & 0.38 & 0.44\\
\hline
\hline
$f_{PS}$
& 0.13 & 0.16 & 0.38 & 0.21 & 0.23 & --& --& 0.42
& & & & &  \\
\hline
$f_V$ & & & & & & & & & 0.20 & 0.21 & 0.41 & 0.20 & 0.27 \\
\hline\hline
\end{tabular}

\vspace{1cm}

\noindent {\bf Table 2}.
The inclusive partial widths of the $\bar{B}^0$ meson
for the different choices
of the continuum threshold $M_X^{(0)}$
(in units $|V_{bc}|^2/|0.039|^2$
$ps^{-1}$). We use the values of the CKM parameters $|V_{sc}|=0.974$, 
$|V_{dc}|=|V_{us}|=0.221$. The CKM matrix
element $|V_{cb}|$ is calculated using the experimental value of
$\Gamma_{tot}(\bar{B}^0)=0.641~ps^{-1}$. Case $a$: $M_X^{(0)}=
M_V+m_{\pi}$, case $b$: $M_X^{(0)}=M_P+m_{\pi}$.\\[3mm]
{\Large
\bc
\vspace{0.5cm}
\begin{tabular}{|c|c|c|c|c|}
\hline\hline
 & & & & \\
Decay mode & $N_{\rm exc}$ & $N_{\rm inc}$ & $a$ & $b$ \\
\hline\hline
$\bar{b}\to \bar{c}+e+\nu_e$ & 2 & 1 & 0.074 & 0.082\\
\hline
$\bar{b}\to \bar{c}+\mu+\nu_{\mu}$ & 2 & 1 & 0.074 & 0.081\\
\hline
$\bar{b}\to \bar{c}+\tau+\nu_{\tau}$ & 2 & 1 & 0.015 & 0.017 \\
\hline
$\bar{b}\to \bar{c}+u+\bar{d}$ & 12 & 6 & 0.313 & 0.358\\
\hline
$\bar{b}\to \bar{c}+c+\bar{s}$ & 12 & 6 & 0.123 & 0.135\\
\hline
$\bar{b}\to \bar{c}+u+\bar{s}$ & 12 & 6 & 0.016 & 0.018\\
\hline
$\bar{b}\to \bar{c}+c+\bar{d}$ & 12 & 6 & 0.006 & 0.006\\
\hline
$\bar{B}^0 \to N\bar{\Lambda}_c,\Lambda_c\bar{\Xi}_c$ & & 2 & 0.023 & 0.023\\
\hline
$\bar{b}\to \bar{u}$ & 54 & 29 & 0.007 & 0.008 \\
\hline\hline
$\Sigma_{tot}$ & 108 & 58 & 0.652 & 0.728\\
\hline\hline
$|V_{bc}|$ & & & 0.0386 & 0.0366\\
\hline
$Br_{SL}$ & & & $11.35\%$ & $11.26\%$\\
\hline
$n_c$ & & & 1.198 & 1.194\\
\hline\hline
\end{tabular}
\ec
}

\newpage

\noindent {\bf Table 3.}
The two--meson branching fractions of $\bar{B}^0$
(in $\%$).\\[5mm]
{\Large 
\bc

\begin{tabular}{|c|c|c|c|c|}
\hline\hline
Decay mode & This work & ref.\cite{NRSX92} & ref.\cite{NS97} &
Exp. data \cite{BHP96}\\ 
\hline Class I & & &\\ \hline\hline $\bar{B}^0\to
D^+\pi^-$ &  $0.34a_1^2$ & $0.257a_1^2$ & $318a_1^2$ &
$0.29\pm0.05$\\ 
\hline $\bar{B}^0\to D^+\rho^-$ &  $0.78a_1^2$ &
$0.643a_1^2$ & $0.778a_1^2$ & $0.81\pm0.17$\\ 
\hline $\bar{B}^0\to
D^{*+}\pi^-$ &  $0.32a_1^2$ & $0.247a_1^2$ & $0.296a_1^2$ &
$0.25\pm0.05$\\ 
\hline $\bar{B}^0\to D^{*+}\rho^-$ &  $0.83a_1^2$ &
$0.727a_1^2$ & $0.870a_1^2$ & $0.70\pm0.16$\\ 
\hline\hline $\bar{B}^0\to
D^+D_s^-$ &  $1.05a_1^2$ & $0.879a_1^2$ & $1.004a_1^2$ &
$0.82\pm0.36$\\ 
\hline $\bar{B}^0\to D^+D_s^{*-}$ &  $0.92a_1^2$ &
$0.817a_1^2$ & $0.830a_1^2$ & $0.95\pm0.45$\\ 
\hline $\bar{B^0}\to
D^{*+}D_s^-$ & $0.66a_1^2$ & $0.597a_1^2$ & $0.603a_1^2$ &
$0.85\pm0.33$\\ 
\hline $\bar{B}^0\to D^{*+}D_s^{*-}$ &  $2.32a_1^2$ &
$2.097a_1^2$ & $2.414a_1^2$ & $1.85\pm0.72$\\ 
\hline $B^0\to D^+K^-$
& $0.028a_1^2$ & $0.020a_1^2$ & $0.037a_1^2$ & \\ 
\hline $B^0\to
D^-K^{*-}$ & $0.045a_1^2$ & $0.035a_1^2$ & $0.041a_1^2$ & \\ 
\hline
$\bar{B}^0\to D^{*+}K^-$ & $0.015a_1^2$ & $0.019a_1^2$ & $0.022a_1^2$ & \\
\hline
$\bar{B}^0\to D^{*+}K^{*-}$ & $0.033a_1^2$ & $0.042a_1^2$ & $0.049a_1^2$ & \\
\hline\hline
Class II & & & &\\
\hline\hline
$B^0\to \pi^0\bar{D}^0$ &  $0.28a_2^2$ & $0.164a_2^2$ & $0.084a_2^2$ & $<0.033$\\
\hline
$\bar{B}^0\to \pi^0\bar{D}^{*0}$ &  $0.21a_2^2$ & $0.230a_2^2$ & $0.116a_2^2$ & $<0.055$\\
\hline
$\bar{B}^0\to \rho^0\bar{D}^0$ &  $0.12a_2^2$ & $0.111a_2^2$ & $0.078a_2^2$ & $<0.055$\\
\hline
$\bar{B}^0\to \rho^0\bar{D}^{*0}$ & $0.21a_2^2$ & $0.240a_2^2$ & $0.199a_2^2$ & $<0.117$\\
\hline\hline
$\bar{B}^0\to K^0+\eta_c$ &  $0.81a_2^2$ & $$ & & \\
\hline
$\bar{B}^0\to K^0J/\psi$ & $0.64a_2^2$ & $2.262a_2^2$ & $0.800a_2^2$ & $0.075\pm0.021$\\
\hline
$\bar{B}^0\to K^{*0}\eta_c$ &  $0.41a_2^2$ & & &\\
\hline
$\bar{B}^0\to K^{*0}J/\psi$ &  $1.46a_2^2$ & $3.645a_2^2$ & $2.518a_2^2$ & $0.151\pm0.091$\\
\hline\hline
\end{tabular}

\end{center}
}
\newpage

\noindent {\bf Table 4.} Form factors for the external $B^+_c$-decays at $q^2=0$. In
parentheses are given the values of corresponding form factors for the
decays of $\bar{B}^0$ or $D^0$.\\[3mm]
\begin{center}
\begin{tabular}{|c|c|c|c|c|}
\hline\hline
 & $B_c \to \eta_c,J/\psi $ & $B_c \to B_s,B_s^*$ & $B_c\to D^0,D^{0*}$ & $B_c\to B^0,B^{0*}$ \\
\hline\hline
$F_1(0)$ & 0.622(0.683) & 0.564(0.780) & 0.089(0.293) & 0.362(0.681) \\
\hline
$A_0(0)$ & 0.477(0.678) & 0.447(0.730) & 0.050(0.214) & 0.282(0.600) \\
\hline
$A_1(0)$ & 0.447(0.623) & 0.453(0.633) & 0.045(0.170) & 0.290(0.502) \\
\hline
$A_2(0)$ & 0.390(0.556) & 0.522(0.464) & 0.040(0.155) & 0.370(0.366) \\
\hline
V(0) & 0.621(0.683) & 2.17(0.777) & 0.073(0.216) & 1.50(0.663) \\
\hline\hline
\end{tabular}
\ec
\vspace{0.3cm}
\begin{center}
\noindent{\bf Table 5.}
Inclusive $B_c$ partial rates (in units of $ps^{-1}$).\\[3mm]
{\Large
\begin{tabular}{|c|c|c|c|c|}
\hline\hline
Decay mode & $N_{\rm exc}$ & $N_{\rm inc}$ & This work & ref.
\cite{BB96}\\
\hline\hline
$\bar{b}\to \bar{c}+e+\nu_e$ & 2 & 1 & 0.061 & 0.075\\
\hline
$\bar{b}\to \bar{c}+\mu+\nu_{\mu}$ & 2 & 1 & 0.061 & 0.075\\
\hline
$\bar{b}\to \bar{c}+\tau+\nu_{\tau}$ & 2 & 1 & 0.013 & 0.018\\
\hline
$\bar{b}\to \bar{c}+u+\bar{d}$ & 12 & 6 & 0.259 & 0.310\\
\hline
$\bar{b}\to \bar{c}+c+\bar{s}$ & 8 & 6 & 0.102 & 0.137 \\
\hline
$\bar{b}\to \bar{c}+u+\bar{s}$ & 12 & 6 & 0.013 & -- \\
\hline
$\bar{b}\to \bar{c}+c+\bar{d}$ & 8 & 6 & 0.006 & -- \\
\hline
$B_c \to \Sigma_c\bar{\Sigma}_c$ & & 2 & 0.011 & --\\
\hline
$\bar{b}\to \bar{u}$ & 46 & 29 & 0.005 &-- \\
\hline\hline
$\Sigma^{\bar{b}}_{tot}$ & 92 & 58 & 0.531 & 0.615\\
\hline\hline
$c\to s+e+\nu_e$ & 2 & 1 & 0.081 & 0.162\\
\hline
$c\to s+\mu+\nu_{\mu}$ & 2 & 1 & 0.077 & 0.162\\
\hline
$c\to s+u+\bar{d}$ & 12 & 6 & 0.698 & 0.905\\
\hline
$c\to s+u+\bar{s}$ & 12 & 6 & 0.023 & --\\
\hline
$c\to d$ & 28 & 14 & 0.036 & -- \\
\hline\hline
$\Sigma^c_{tot}$ & 56 & 28 & 0.915 & 1.229\\
\hline\hline
$\bar{b}c\to \tau+\nu_{\tau}$ & & & 0.052 & 0.056\\
\hline
$\bar{b}c\to c+\bar{s}$ & & & 0.137 & 0.138\\
\hline
PI & & &  $\approx -0.100$ & -0.124\\
\hline\hline
$\Sigma_{tot}$ & 148 & 86 & 1.535 & 1.914\\
\hline\hline
\end{tabular}}
\end{center}

\newpage

\noindent {\bf Table 6. } SL and NL external partial rates for $\bar{b}\to \bar
c$ and $c\to s$ transitions (in units $10^{-3}ps^{-1}$). The values of the 
CKM parameters used in the calculations are the same as in Table 2. 
The values of the QCD two--meson amplitudes are $a_1=1.0$, $a_2=-0.3$ and 
$a_1=c_1(m_c)$, $a_2=c_2(m_c)$ 
for the $\bar{b} \to \bar{c}$ and $c \to s$ transitions, respectively. \\[5mm]
\begin{center}
{\Large
\begin{tabular}{|c|c|c|c|c|c|c|}
\hline\hline
& \multicolumn{3}{|c|}{$\bar{b}\to \bar{c}l\nu_l$}&\multicolumn{3}{|c|}{$c\to sl\nu_l$}\\
\cline{2-7}
& $\eta_c$ & $J/\psi$ & $X_{c\bar{c}}$ & $B_s$ & $B_s^*$ & $X_{\bar{b}s}$\\
\hline\hline
$e\nu_e$ & 13.05 & 26.6 &  20.6 & 22.0 & 51.2 & 7.35\\
\hline
$\mu\nu_{\mu}$ & 13.02 & 26.5 & 20.6 & 21.1 & 48.2 & 6.34\\
\hline
$\tau\nu_{\tau}$ & 4.37 & 7.53 & 1.81 & -- & -- & --\\
\hline\hline
\end{tabular}

\vspace{5mm}

\begin{tabular}{|c|c|c|c|c|c|c|}
\hline\hline
& \multicolumn{3}{|c|}{$\bar{b}\to \bar{c}u\bar{d}$} & \multicolumn{3}{|c|}{$\bar{b}\to \bar{c}u\bar{s}$}\\
\cline{2-7}
    & $\pi^+$ & $\rho^+$ & $X_{u\bar{d}}$ & $K^+$ & $K^{*+}$ & $X_{u\bar{s}}$\\
\hline\hline
$\eta_c$ & 2.25 & 5.15 & 46.4 & 0.178 & 0.289 & 2.30\\
\hline
$J/\psi$ & 1.27 & 3.57 & 95.9 & 0.096 & 0.213 & 4.61\\
\hline
$X_{c\bar{c}}$ & 5.71 & 13.3 & 65.9 & 0.329 & 0.751 & 2.99\\
\hline\hline
\end{tabular}

\vspace{5mm}

\begin{tabular}{|c|c|c|c|c|c|c|c|}
\hline\hline
&\multicolumn{3}{|c|}{$\bar{b}\to \bar{c}c\bar{s}$} &\multicolumn{3}{|c|}{$\bar{b}\to \bar{c}c\bar{d}$}\\
\cline{2-7}
    & $D_s$ & $D_s^*$ & $X_{c\bar{s}}$ & $D^+$ & $D^{*+}$ & $X_{c\bar{d}}$\\
\hline\hline
$\eta_c$ & 8.03 & 6.68 & 10.8 & 0.344 & 0.198 & 0.690\\
\hline
$J/\psi$ & 3.28 & 13.2 & 17.5 & 0.146 & 0.354 & 1.16\\
\hline
$X_{c\bar{c}}$ & 6.79 & 14.0 & 2.68 & 0.426 & 0.440 & 0.245\\
\hline\hline
\end{tabular}

\vspace{5mm}

\begin{tabular}{|c|c|c|c|c|c|c|}
\hline\hline
& \multicolumn{3}{|c|}{$c\to su\bar{d}$} & \multicolumn{3}{|c|}{$c\to s\bar{s}u$}\\
\cline{2-7}
    & $\pi^+$ & $\rho^+$ & $X_{u\bar{d}}$ & $K^+$ & $K^{*+}$ & $X_{u\bar{s}}$\\
\hline\hline
$B_s$ & 96.4 & 65.4 & 0.180 & 8.06 & 0.432 & --\\
\hline
$B_s^*$ & 55.0 & 340 & -- & 3.22 & 0.213 & --\\
\hline
$X_{\bar{b}s}$ & 54.5 & -- & -- & 1.56 & -- & --\\
\hline\hline

\end{tabular}}
\end{center}
\newpage
\begin{center}
{\bf Table 7.}
Two-meson $\bar{b}\to \bar{c}$ rates of $B^+_c$ (in
$10^{-3}ps^{-1}$).\\[5mm] {\Large \begin{tabular}{|c|c|c|c|} \hline
Decay mode & This work & ref.\cite{LM91} & ref.\cite{CC94} \\
\hline\hline
Class I & & &\\
\hline\hline
$B^+_c\to \eta_c\pi^+$ & $2.23a_1^2$ & $1.83a_1^2$ & $3.14a_1^2$ \\
\hline
$B^+_c\to \eta_c\rho^+$ & $5.09a_1^2$ & $4.32a_1^2$ & $8.33a_1^2$\\
\hline
$B^+_c\to J/\psi\pi^+$ & $1.25a_1^2$ & $1.92a_1^2$ & $3.00a_1^2$ \\
\hline
$B^+_c\to J/\psi\rho^+$ & $3.53a_1^2$ & $5.42a_1^2$ & $9.04a_1^2$\\
\hline\hline
$B^+_c\to \eta_cK^+$ & $0.23a_1^2$ & $0.14a_1^2$ & $0.245a_1^2$\\
\hline
$B^+_c\to \eta_cK^{*+}$ & $0.36a_1^2$ & $0.22a_1^2$ & $0.435a_1^2$\\
\hline
$B^+_c\to J/\psi K^+$ & $0.12a_1^2$ & $0.14a_1^2$ & $0.231a_1^2$ \\
\hline
$B^+_c\to J/\psi K^{*+}$ & $0.27a_1^2$ & $0.28a_1^2$ & $0.492a_1^2$\\
\hline\hline
Class II & & & \\
\hline\hline
$B^+_c\to D^+\bar{D}^0$ & $4.13a_2^2$ & & $1.01a_2^2$\\
\hline
$B^+_c\to D^+\bar{D}^{*0}$ & $3.19a_2^2$ & & $1.06a_2^2$\\
\hline
$B^+_c\to D^{*+}\bar{D}^0$ & $1.31a_2^2$ & & $0.992a_2^2$\\
\hline
$B^+_c\to D^{*+}\bar{D}^{*0}$ & $2.01a_2^2$ & & $1.64a_2^2$\\
\hline\hline
Class III & & & \\
\hline\hline
$B^+_c\to \eta_cD_s^+$ & $(3.19a_1+4.19a_2)^2$ & & $(1.39a_1+2.44a_2)^2$\\
\hline
$B^+_c\to \eta_cD_s^{*+}$ & $(2.91a_1+2.45a_2)^2$ & & $(1.28a_1+2.34a_2)^2$\\
\hline
$B^+_c\to J/\psi D_s^+$ & $(2.04a_1+3.60a_2)^2$ & & $(1.26a_1+2.40a_2)^2$\\
\hline
$B^+_c\to J/\psi D_s^{*+}$ & $(4.09a_1+4.79a_2)^2$ & & \\
\hline\hline
$B^+_c\to \eta_cD^+$ & $(0.58a_1+0.90a_2)^2$ & & $(0.24a_1+0.54a_2)^2$\\
\hline
$B^+_c\to \eta_cD^{*+}$ & $0.45a_1+0.81a_2)^2$ & & $(0.22a_1+0.53a_2)^2$\\
\hline
$B^+_c\to J/ \psi D^+$ & $(0.37a_1+0.54a_2)^2$ & & $(0.22a_1+0.54a_2)^2$\\
\hline
$B^+_c\to J/\psi D^{*+}$ & $(0.59a_1+0.98a_2)^2$ & &\\
\hline\hline
\end{tabular}}
\end{center}

\newpage
\begin{center}
\vspace{1cm}
{\bf Table 8.} Two-meson $c\to s$ rates of $B^+_c$ (in
$10^{-3}ps^{-1}$).\\[5mm] {\Large \begin{tabular}{|c|c|c|c|c|}
\hline\hline
Decay Mode & This work & ref.\cite{LM91} & ref.\cite{LM91} &
ref.\cite{CC94} \\ \hline\hline Class I & & &\\ \hline\hline
$B^+_c\to B_s\pi^+$ & $52.86a_1^2$ & $47.2a_1^2$ & $66.8a_1^2$ &
$88.7a_1^2$\\ \hline $B^+_c\to B_s\rho^+$ & $35.88a_1^2$ &
$19.0a_1^2$ & $30.7a_1^2$ & $68.1a_1^2$\\ \hline $B^+_c\to
B_s^*\pi^+$ & $30.14a_1^2$ & $38.9a_1^2$ & $52.7a_1^2$ &
$78.4a_1^2$\\ \hline $B^+_c\to B_s^*\rho^+$ & $186.46a_1^2$ &
$175.6a_1^2$ & $231.1a_1^2$ & $228a_1^2$\\ \hline $B^+_c\to B_0
\pi^+$ & $2.30a_1^2$ & $1.47a_1^2$ & $2.87a_1^2$ & $5.01a_1^2$\\
\hline
$B^+_c\to B_0 \rho^+$ & $2.93a_1^2$ & $1.43a_1^2$ & $3.25a_1^2$ & $9.07a_1^2$\\
\hline
$B^+_c\to B_0^*\pi^+$ & $1.19a_1^2$ & $2.40a_1^2$ & $1.94a_1^2$ & $4.41a_1^2$\\
\hline
$B^+_c\to B_0^*\rho^+$ & $10.3a_1^2$ & $13.4a_1^2$ & $13.5a_1^2$ & $18.07a_1^2$\\
\hline\hline
Class II & & &\\
\hline\hline
$B^+_c\to B^+\bar{K}^0$ & $36.54a_2^2$ & $42.8a_2^2$ & $93.3a_2^2$ & $147a_2^2$\\
\hline
$B^+_c\to B^+\bar{K}^{*0}$ & $20.93a_2^2$ & $15.2a_2^2$ & $36.6a_2^2$ & $104a_2^2$\\
\hline
$B^+_c\to B^{*+}\bar{K}^0$ & $13.55a_2^2$ & $47.1a_2^2$ & $43.0a_2^2$ & $111a_2^2$\\
\hline
$B^+_c\to B^{*+}\bar{K}^{*0}$ & $125.10a_2^2$ & $223.5a_2^2$ & $248.9a_2^2$ & $214a_2^2$\\
\hline
$B^+_c\to B^+\pi^0$ & $1.56a_2^2$ & $0.73a_2^2$ & $1.44a_2^2$ & $2.51a_2^2$\\
\hline
$B^+_c\to B^+\rho^0$ & $1.95a_2^2$ & $0.71a_2^2$ & $1.63a_2^2$ & $4.53a_2^2$\\
\hline
$B^+_c\to B^{*+}\pi^0$ & $0.80a_2^2$ & $1.20a_2^2$ & $0.97a_2^2$ & $2.20a_2^2$\\
\hline
$B^+_c\to B^{*+}\rho^0$ & $6.93a_2^2$ & $6.70a_2^2$ & $6.73a_2^2$ & $9.05a_2^2$\\
\hline\hline
\end{tabular}}
\ec
\vspace{2cm}

\newpage
\vspace{0.5cm}
\begin{figure}[htb]
\centerline{\epsfxsize=7.0in \epsfysize=8.0in \epsffile{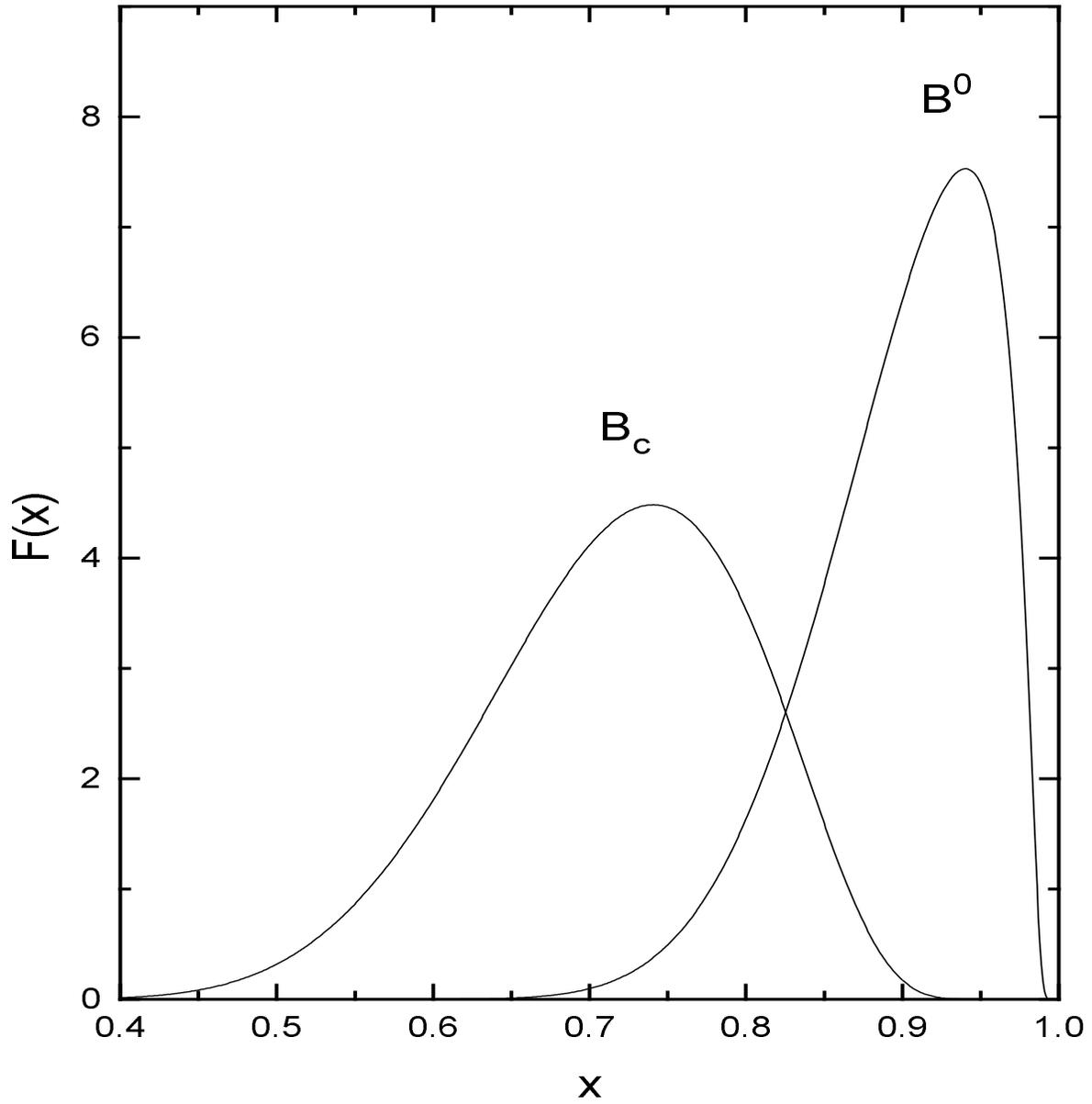}}
\caption{Distribution functions $F_{\bar{b}}(x)$ for $B^0$  and $B_c$
mesons.}
\end{figure}

\end{document}